\documentclass[12pt]{article}
\usepackage{graphicx}

\newcommand\pubnumber{SNSN-323-63}
\newcommand\pubdate{\today}

\def\institute{Austrian Acedemy Of Sciences\\
Nikolsdorfer Gasse 18, 1050 Vienna, Austria}
\def\behalf{\footnote{on behalf of the CMS Collaboration}}
\def\Title#1{\begin{center} {\Large #1 } \end{center}}
\def\Author#1{\begin{center}{ \sc #1} \end{center}}
\def\Address#1{\begin{center}{ \it #1} \end{center}}

\newcommand\pubblock{\rightline{\begin{tabular}{l} \pubnumber\\
         \pubdate  \end{tabular}}}
\newenvironment{Abstract}{\begin{quotation}  }{\end{quotation}}
\newenvironment{Presented}{\begin{quotation} \begin{center} 
             PRESENTED AT\end{center}\bigskip 
      \begin{center}\begin{large}}{\end{large}\end{center} \end{quotation}}

\def\beq{\begin{equation}}
\def\eeq#1{\label{#1}\end{equation}}
\def\eeqn{\end{equation}}

\def\beqa{\begin{eqnarray}}
\def\eeqa#1{\label{#1}\end{eqnarray}}
\def\eeqan{\end{eqnarray}}

\let\bar=\overbar

\def\Dslash{\not{\hbox{\kern-4pt $D$}}}
\def\dslash{\not{\hbox{\kern-2pt $\del$}}}

\def\msb{{\bar{\ssstyle M \kern -1pt S}}}

\begin{document}
\begin{titlepage}
\pubblock

\vfill
\Title{Studies of top quark properties in CMS}
\vfill
\Author{ Dennis Schwarz\behalf }
\Address{\institute}
\vfill
\begin{Abstract}
In this article, recent studies of top quark properties in CMS are presented. The discussed analyses are measurements of the charge asymmetry, CP violation and jet mass, all carried out in final states of $\textrm{t}\bar{\textrm{t}}$ production with one or two leptons. The data were recorded with the CMS detector in the years 2016 to 2018.
\end{Abstract}
\vfill
\begin{Presented}
$15^\mathrm{th}$ International Workshop on Top Quark Physics\\
Durham, UK, 4--9 September, 2022
\end{Presented}
\vfill
\end{titlepage}
\def\thefootnote{\fnsymbol{footnote}}
\setcounter{footnote}{0}

\section{Introduction}
Weighing in at about $170\;\textrm{GeV}$, the top quark is the heaviest particle in the standard model~(SM) and has a Yukawa coupling of about unity. Thus, it plays a special role in the electroweak sector of the SM. Beyond the SM~(BSM) physics that may exist at energy scales above the direct reach of the LHC could manifest in subtle deviations from SM predictions of top quark properties. In this article, a selection of recent analyses performed by the CMS Collaboration is presented that target precise measurements of top quark properties. A detailed description of the CMS experiment can be found in Ref.~\cite{cms}.

\section{Measurement of the charge asymmetry in $\textrm{t}\bar{\textrm{t}}$ production}
At the LHC, pairs of top quarks and top antiquarks~($\textrm{t}\bar{\textrm{t}}$) are abundantly produced. In Ref.~\cite{asymm} the charge asymmetry of top quarks and top antiquarks is accessed by measuring the central-forward asymmetry. It is defined as
\begin{equation}
A_\textrm{C} = \frac{N(\Delta|y|>0)-N(\Delta|y|<0)}{N(\Delta|y|>0)+N(\Delta|y|<0)},
\end{equation}
where $\Delta|y|$ is the difference of the absolute values of the rapidities of top quark and top antiquark and $N(\Delta|y|>0)$ ($N(\Delta|y|<0)$) corresponds to the number of observed events where this difference is positive (negative). In the SM at leading order, $A_\textrm{C}$ is predicted to be zero. At higher orders contributions from $\textrm{q}\bar{\textrm{q}}\rightarrow\textrm{t}\bar{\textrm{t}}$ production increase the expected value to about $1\%$. However, BSM physics may shift the asymmetry to even larger values. 

\begin{figure}[htb]
	\centering
	\includegraphics[width=.49\textwidth]{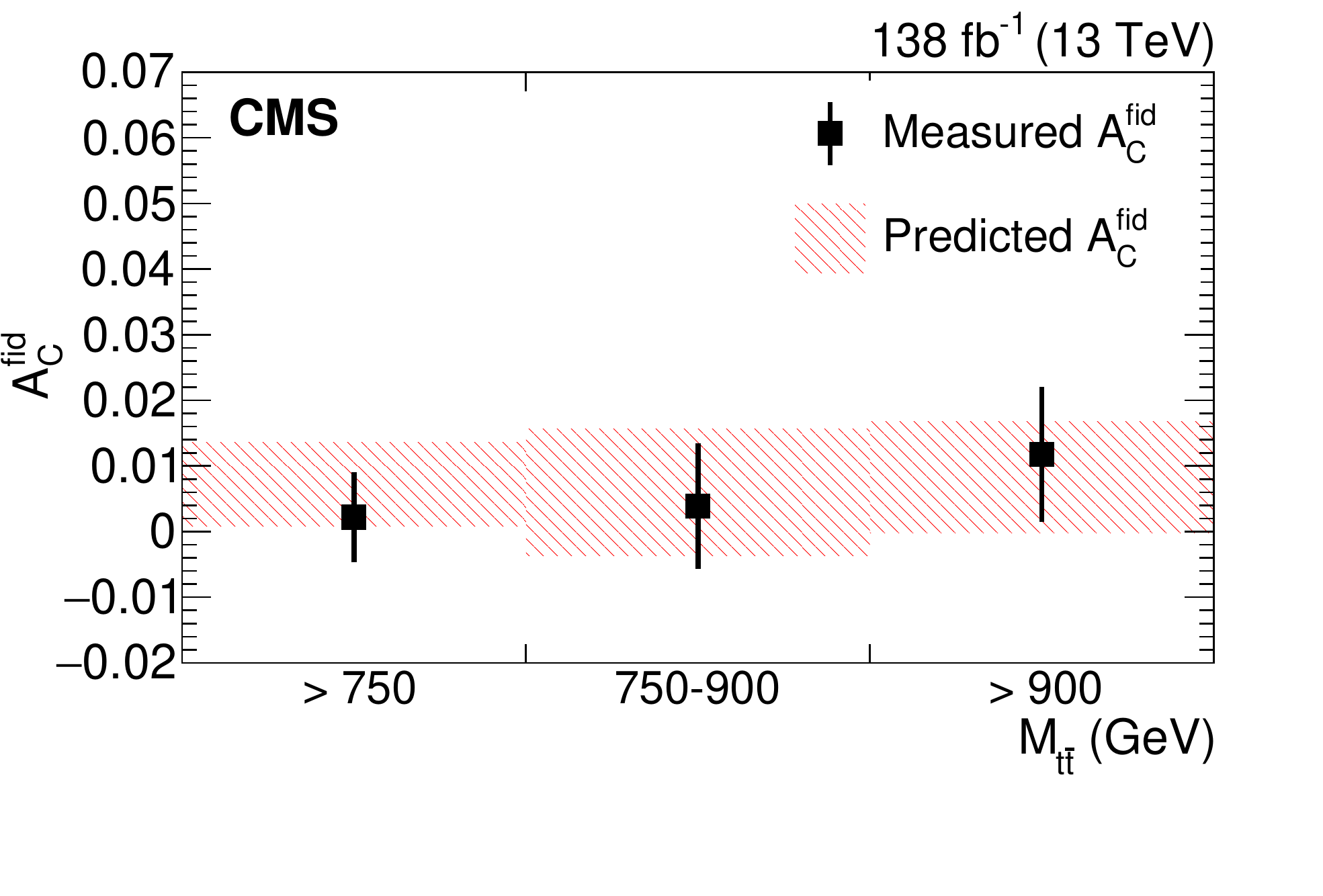}
	\includegraphics[width=.49\textwidth]{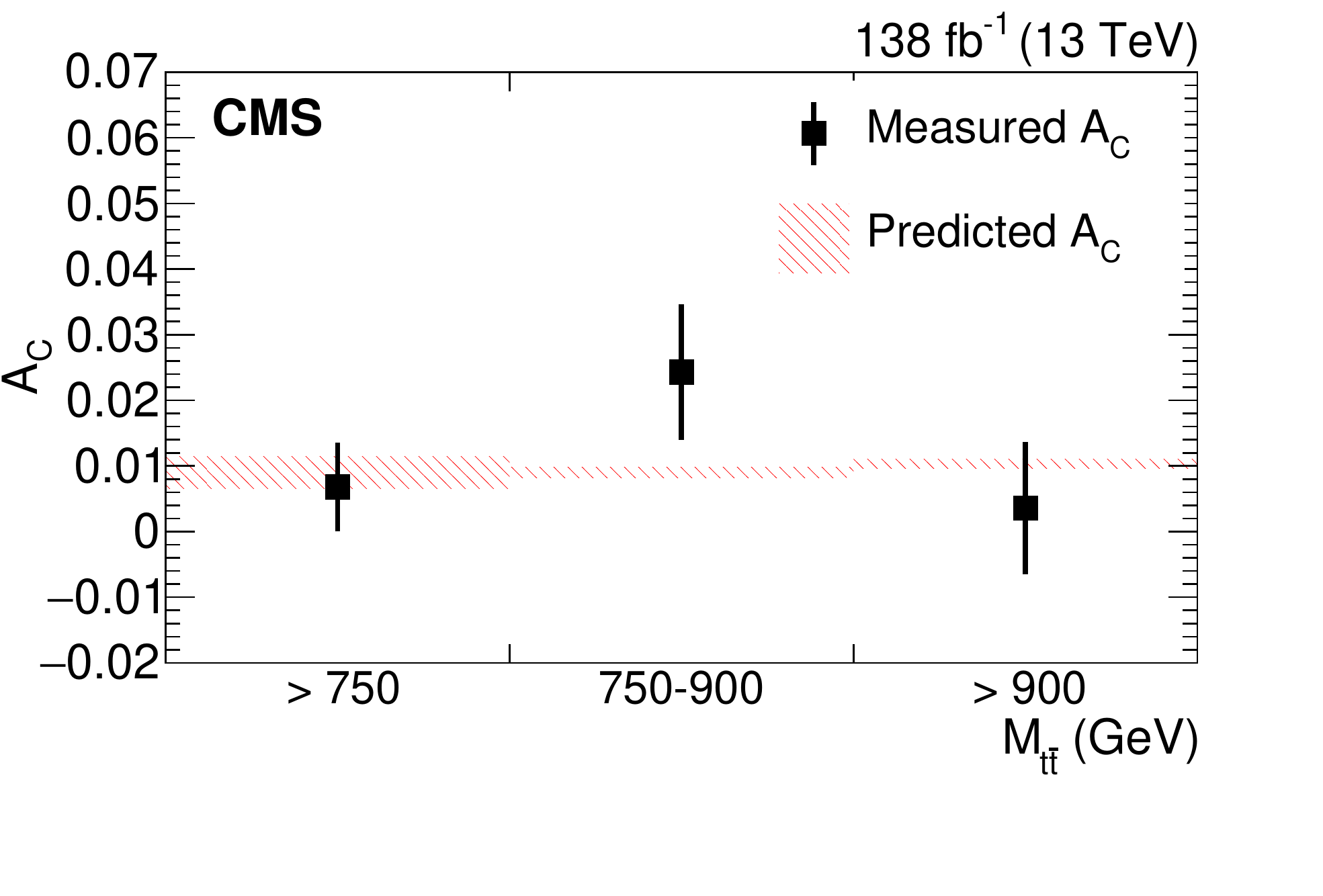}
	\caption{Charge asymmetry in the fiducial~(left) and full~(right) phase space of $\textrm{t}\bar{\textrm{t}}$ production. The data~(markers) are compared to the SM prediction~(colored area). The total uncertainty in the measurement is shown by vertical error bars and the uncertainty in the SM prediction is indicated by the size of the colored area. Published in Ref.~\cite{asymm}.}
	\label{fig:asymm}
\end{figure}

In the measurement, top quarks with high momenta are selected in order to enrich the analyzed data in $\textrm{q}\bar{\textrm{q}}\rightarrow\textrm{t}\bar{\textrm{t}}$ production. Furthermore, the lepton+jets channel of $\textrm{t}\bar{\textrm{t}}$ is selected, where the hadronic top quark decay is reconstructed either with a large-$R$ top-tagged jet, a large-$R$ W-tagged jet and a small-$R$ jet, or a combination of three small-$R$ jets, where $R$ is the distance parameter used in the jet clustering that is set to either 0.4 or 0.8. After reconstructing the full $\textrm{t}\bar{\textrm{t}}$ system, the central-forward asymmetry is measured as a function of the invariant mass of the $\textrm{t}\bar{\textrm{t}}$ system. The result is shown in Fig.~\ref{fig:asymm} for the fiducial and full phase space. The measurement is in agreement with the prediction of the SM. The dominating uncertainties arise from QCD scale variations in the theoretical prediction, the $\textrm{t}\bar{\textrm{t}}$ modeling in simulation and the jet energy calibration.

\section{Measurement of the CP violation in $\textrm{t}\bar{\textrm{t}}$ production}
\begin{figure}[htb]
	\centering
	\includegraphics[width=.49\textwidth]{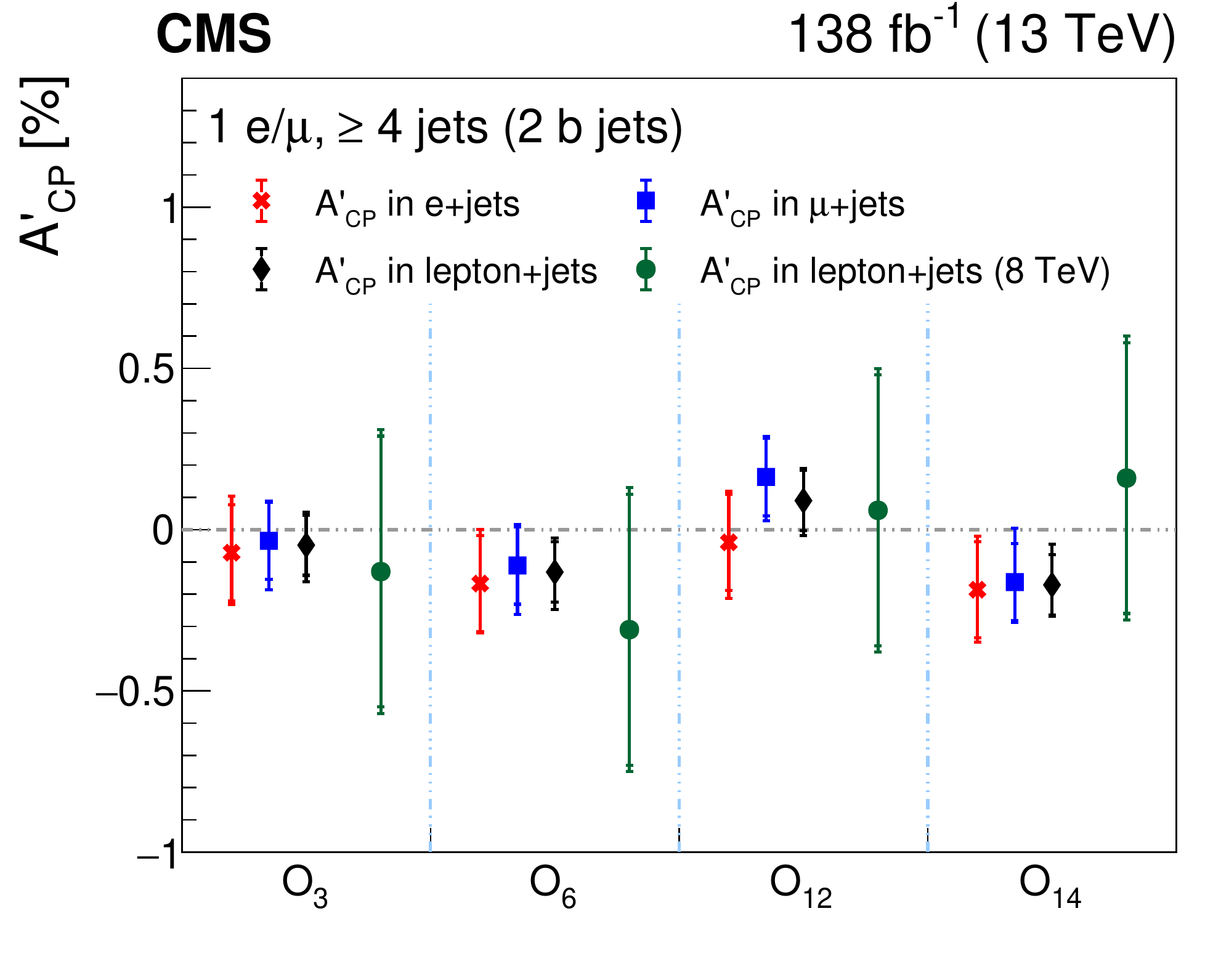}
	\caption{Asymmetry in CP-observables for the e+jets~(red markers), $\mu$+jets~(blue markers) and combined~(black markers) channels. The results are compared to a previous analysis~(green markers). The total uncertainty is shown by the vertical error bars. Published in Ref.~\cite{cp}.}
	\label{fig:cp}
\end{figure}

In Refs.~\cite{cp} and~\cite{cpdilep} the $\textrm{t}\bar{\textrm{t}}$ production is tested for CP-violation. In both analyses CP-observables are defined that are strictly symmetric in the SM. The CP-observables are constructed from the 4-momenta of the final state objects in $\textrm{t}\bar{\textrm{t}}$ production - namely leptons, missing momentum, b-tagged jets and light-flavor jets. Thus, similar to the measurement of $A_\textrm{C}$ presented above, the full $\textrm{t}\bar{\textrm{t}}$ system needs to be reconstructed. Then, the asymmetry for each CP-observable $O_i$ is defined as
\begin{equation}
	A_{i,\textrm{CP}} = \frac{N(O_i>0)-N(O_i<0)}{N(O_i>0)+N(O_i<0)},
\end{equation}
where $N(O_i>0)$ and $N(O_i<0)$ are the number of observed events with positive and negative values of $O_i$, respectively. Figure~\ref{fig:cp} shows the measurement of $A_{i,\textrm{CP}}$ for the four CP-observables that are reconstructed in Ref.~\cite{cp}. The results are displayed separately by lepton channel and compared to an earlier measurement, showing the consistency between lepton flavors and improvement in precision. Overall, the measurement is consistent with the SM expectation of $A_{i,\textrm{CP}}=0$ for all CP-observables. The dominant experimental uncertainties in this measurement are connected to the calibration of the jet energy scale and resolution.

In the dilepton channel of $\textrm{t}\bar{\textrm{t}}$ production, a special reconstruction is needed because two final state neutrinos escape the detector but only one measurement of missing transverse momentum is obtained. Otherwise, the search for CP-violation in the dilepton channel~\cite{cpdilep} is performed similarly and is in agreement with the SM prediction as well. Both analyses are furthermore interpreted in the scope of BSM models, which is detailed in Refs.~\cite{cp,cpdilep}.

\section{Measurement of the jet mass in hadronic decays of boosted top quarks}
\begin{figure}[htb]
\centering
\includegraphics[width=.49\textwidth]{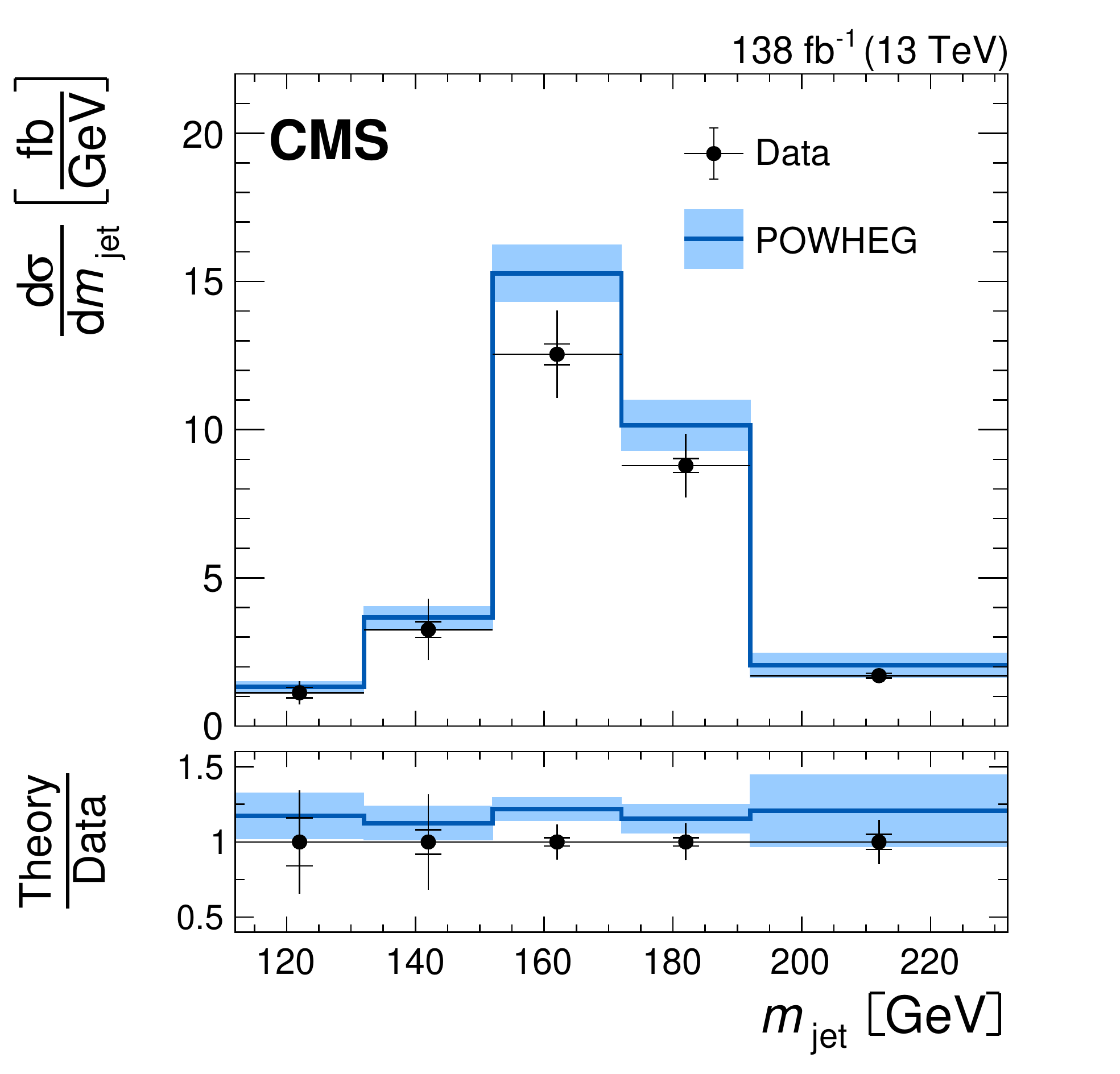}
\includegraphics[width=.49\textwidth]{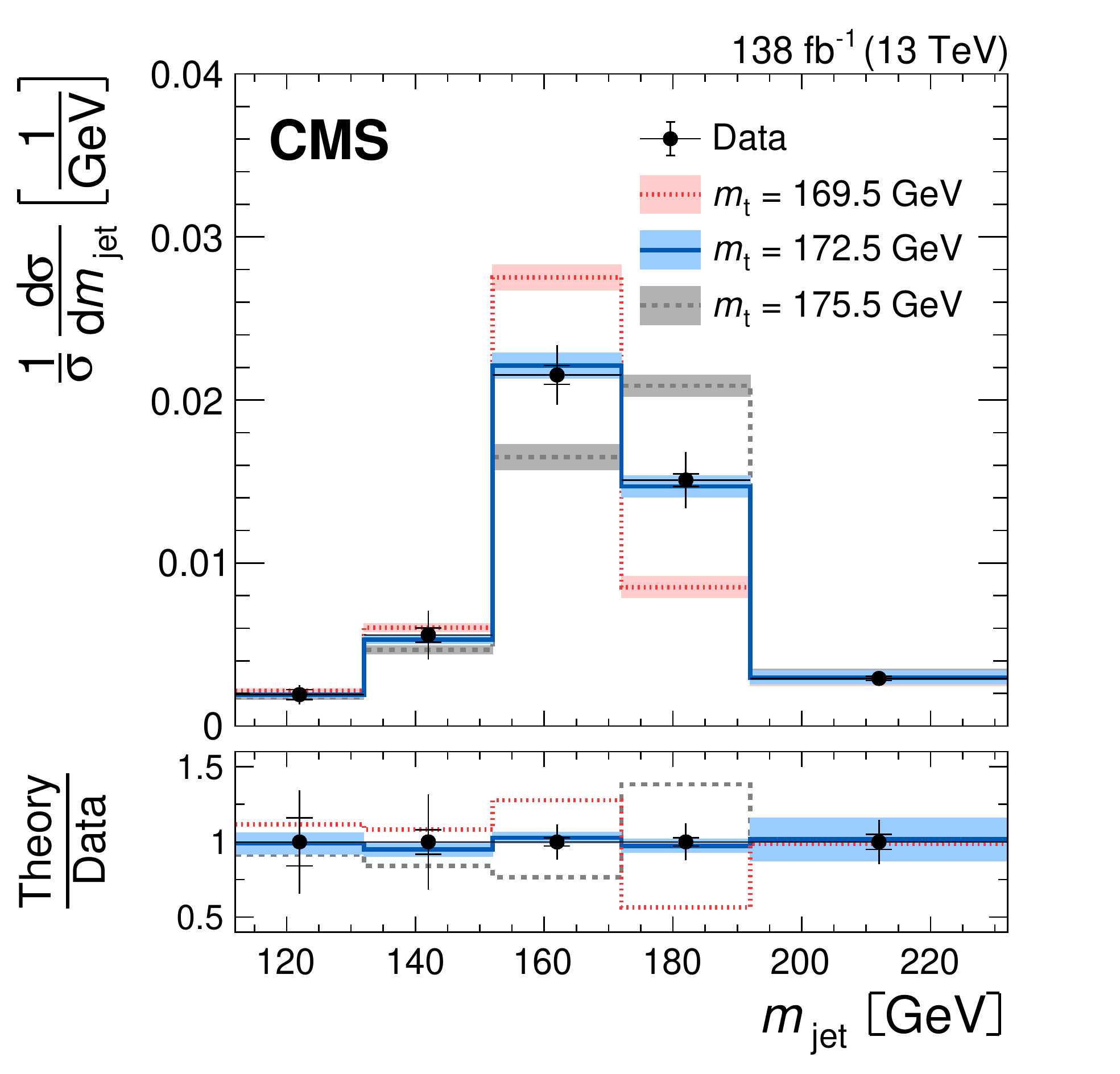}
\caption{Differential $\textrm{t}\bar{\textrm{t}}$ production cross section as a function of the jet mass~(left). On the right, all distributions are normalized. Data~(markers) are compared to the prediction from simulation for various values of the top quark mass. The vertical error bars include statistical~(inner) and total uncertainties~(outer). Theoretical uncertainties are drawn as colored areas. Published in Ref.~\cite{mjet}.}
\label{fig:mjet}
\end{figure}

In Ref.~\cite{mjet} the $\textrm{t}\bar{\textrm{t}}$ production cross section is measured as a function of the jet mass. The jet mass is the invariant mass of the sum of all 4-momenta of jet constituents and is thus sensitive to the mass of the particle that initiated the jet. It is widely used to identify large-$R$ jets that originate from hadronic decays of boosted W/Z/H bosons or top quarks. Also, this jet substructure observable is an important test to our understanding of quantum chromodynamic~(QCD) processes at large and small energy scales. In this measurement, the lepton+jets channel of $\textrm{t}\bar{\textrm{t}}$ is selected. The hadronically decaying top quark is reconstructed with a single large-$R$ jet using the XCone algorithm~\cite{xcone} and a subsequent clustering of small-$R$ subjets, which acts similar to a trimming algorithm~\cite{trimming}. With respect to an earlier measurement~\cite{mjet16}, the precision is increased by calibrating the jet mass scale using the reconstructed W boson mass and tuning the final state radiation modeling in simulation using the N-subjettiness~\cite{nsub} ratio $\tau_3/\tau_2$. Figure~\ref{fig:mjet} shows the unfolded differential $\textrm{t}\bar{\textrm{t}}$ production cross section as a function of the jet mass. The normalized distribution is sensitive to the top quark mass and is used to extract this fundamental parameter. With a measured value of $m_\textrm{t} = 172.76 \pm 0.81\;\textrm{GeV}$ the accuracy is close to the most precise measurements of $m_\textrm{t}$ at much lower top quark energies. Despite largely reduced jet energy scale uncertainties, jet-related uncertainties still dominate the measurement in terms of experimental uncertainties. However, many experimental, modeling and theoretical uncertainties are now on the same level such that further calibrations of single uncertainty sources will not drastically increase the precision of this measurement. 

\section{Summary}
The CMS Collaboration has an extensive program of measurements of various top quark properties. Many analyses reach high precision by performing dedicated calibrations and reconstruction techniques in order to reduce experimental uncertainties and stress-test the SM. With the data available from the LHC, even precise measurements at high top quark momenta are feasible and the SM can be tested in a large phase space. However, all measurements presented in this article are consistent with the SM predictions. 

\pagebreak


\end{document}